\documentclass[onecolumn,dvipdfmx]{article}
\usepackage[utf8]{inputenc}
\usepackage{amsmath}
\usepackage{txfonts}
\usepackage{graphicx}
\title{$\zeta$-Pad{\'e} SRWS theory with lowest order approximation}

\author{Yoshiki Ueoka \thanks{u.yoshiki.phys@gmail.com} \thanks{independent}\\
Osaka Japan}

\begin{document}
\maketitle

\begin{abstract}
In my previous preprint about SRWS-$\zeta$ theory[Y.Ueoka,viXra:2205.014,2022], I proposed an approximation of rough averaged summation  of typical critical Green function for the Anderson transition in the Orthogonal class. 
In this paper, I remove a rough approximate summation for the series of the typical critical Green function by replacing summation with integral.
Pad{\'e} approximant is used to take a summation. 
The perturbation series of the critical exponent $\nu$ of localization length from upper critical dimension is obtained. 
The dimensional dependence of the critical exponent is again directly related with Riemann $\zeta$ function. 
Degree of freedom about lower critical exponent improve estimate compared with previous studies. 
When I fix lower critical dimension equal to two, I obtained similar estimate of the critical exponent compared with fitting curve estimate of the critical exponent[E.Tarquini et al.,PhysRevB.95(2017)094204].
\end{abstract}
\newpage


\section{Introduction}
The Anderson transition is a disorder driven quantum phase transition exhibiting critical phenomena.
The critical exponent is thought to depend only on fundamental properties of the system such as dimensionality and symmetry.
In my previous preprints\cite{ueoka2022a,ueoka2022b}, I propose the new theoretical framework(SRWS or SRWS-$\zeta$) to understand the Anderson transition in the Orthogonal class.
The key points is power series expansion of the typical critical Green function and its approximate summation method.
In this paper, I use better approximation for power series than my previous paper and take a summation by lowest order.
Then, typical critical Green function is obtained closed form.
We can extract the critical exponent $\nu$.
The explicit expression of the dimensional dependence of $\nu$ tells us the number theory is deeply related with the critical phenomena.
A degree of freedom about lower critical dimension appears in high dimensional approximation.
By using high dimensional approximation, I get better estimate of the critical exponent compared with already known numerical estimate.

\section{Anderson transition and SRWS Green function}

In this section, I review starting equation used later in this paper.
The relevant dimensionality is explicitly given by a spectral dimension of a lattice in SRWS.

The typical critical Green function of SRWS theory is given by,
\begin{eqnarray}
    	G_{\mathrm{typical}}(|x-y|,z) &\simeq&  z^{|x-y|+1} \sum_{n=0}^{\infty} c_n z^n  \\ 
	c_n &=&\sum_{t=|x-y|}^{|x-y|+n}A_{xy}(t)\left(\frac{2}{W}\right)^{t+1}\binom{n+|x-y|}{t} \\
	A_xy(t) & =&  \left \{
		\begin{array}{l}
		k^t\left( \frac{t^2}{l} \right)^{-dt/2l} \;\;  (t<l) \\
		k^t t^{-d/2} \;\; (t\ge l)
   		\end{array} \right.  \\
	l&=&\frac{|x-y|^2}{4D}
\end{eqnarray}

\section{the critical exponent obtained from SRWS with approximations}
The treatment of binomial coefficient is a little bit difficult. 
Here, I approximate by using only $t\ge l$ terms.First, I prepare a approximate formula used later,
\begin{equation}
\sum_{t=|x-y|}^{|x-y|+n} (2k/W)^t\binom{n+|x-y|}{t}\simeq\sum_{t=0}^{|x-y|+n} (2k/W)^t\binom{n+|x-y|}{t}=(2k/W)^{n+|x-y|}
\end{equation}
By approximating summation by integration,we get following approximate formula
\begin{equation}
\int_{|x-y|}^{|x-y|+n} (2k/W)^t\binom{n+|x-y|}{t}\mathrm{d}t\simeq (2k/W)^{n+|x-y|} 
\end{equation}
Therefore, I approximate $c_n$ as
\begin{eqnarray}
c_n&\simeq&\frac{1}{W} \int_{t=|x-y|}^{|x-y|+n}t^{-d/2}\left( \frac{2k}{W} \right)^t \binom{n+|x-y|}{t} \nonumber \\
&=&\left[ t^{-d/2}(2k/W)^t\right]_{t=|x-y|}^{t=n+|x-y|}+(d/2)\int_{t=|x-y|}^{|x-y|+n}t^{-d/2}\left( \frac{2k}{W} \right)^t \nonumber \\
&\simeq&\left[ t^{-d/2}(2k/W)^t\right]_{t=|x-y|}^{t=n+|x-y|}+(d/2)\sum_{t=|x-y|}^{|x-y|+n}t^{-d/2}\left( \frac{2k}{W} \right)^t \nonumber \\
\end{eqnarray}
By remaining $n$-dependent terms only,
\begin{equation}
c_n\simeq \frac{1}{W} ((n+|x-y|)^{-d/2}a^{n+|x-y|})+(d/2)\Phi(a,d/2+1,n+|x-y|+1)
\end{equation}
Here,
\begin{equation}
a=\frac{2k}{W}
\end{equation}
Considering summation over $n$,first term of $c_n$ becomes function of $za$. 
Therefore, contribution from the first term gives the term which $a$-dependence is proportional to $a$.
So, We can ignore first term of $c_n$ and remain second term of $c_n$.
Then, we get,
\begin{equation}
G_{\mathrm{typical}} (|x-y|,z)\simeq (z^{|x-y|+1} d)/k a\sum_{n=0}^{\infty} \Phi(a,d/2+1,n+|x-y|+1)  z^n
\end{equation}
Here,
\begin{eqnarray}
\Phi(a,d/2+1,n+|x-y|+1)&=&(1+n+z)^{-d/2-1}+\sum_{k=0}^{\infty}(-1)^k \frac
{(1+n+|x-y|)^k}{k!} \nonumber \\
&\;&(1+d/2)_k Li_{d/2+k+1} (a)
\end{eqnarray}
I ignore first term which does not depend on $a$, and take lowest order cut-off by degree of $k=1$, 
\begin{equation}
\Phi(a,d/2+1,n+|x-y|+1)\simeq\sum_{k=0}^{1}(-1)^k \frac
{(1+n+|x-y|)^k}{k!} (1+d/2)_k Li_{d/2+k+1} (a)
\end{equation}
Then, I take Pad{\`e} approximant of order $[0/1],[0/2],\cdots$ about variable $n$ and take limit$|x-y|\rightarrow \infty$.From observation of behavior of Pad{\`e} approximant and again take limit of $z\rightarrow\infty$. For $k\ge 1$
\begin{equation}
\sum_{n=0}^{\infty}(1+n+|x-y|)^k z^n\simeq \frac{1}{k}|x-y|^k z^{-|x-y|}
\end{equation}
At high dimensional limit $a\rightarrow 0$,for positive $c$ and real number $c_l$,
\begin{equation}
Li_{d/2+k+1} (a)\simeq Li_{d/2+c_l+1} (a)\simeq a
\end{equation} 
By summarizing calculation above, we get with constant $c$ as a degree of freedom.
\begin{equation}
G_{\mathrm{typical}}(|x-y|,z)\propto 1+|x-y|^k(d/2+1)\frac{Li_{d/2+c_l+1}(a)}{a}
\end{equation} 
By taking logarithm, we can extract localization length by,
\begin{equation}
\xi\simeq -\frac{4D\ln{G_{\mathrm{typical}}(|x-y|,z)/z}}{|x-y|^2}
\end{equation}
I take Pad{\'e} approximant about variable $|x-y|$ of order $[2/0] or [3/1],[4/2]$(all approximation gives same final result), and remove $|x-y|^2 $ dependence.
Then series expansion at $a=1$ up to 1st degree, we obtain
\begin{equation}
\xi=\xi_0(d)+\xi_1(d)(a-1)=\xi_0(d)\left(1+\frac{\xi_1(d)}{\xi_0(d)}(a-1)\right)
\end{equation}
Then dimensional dependent term of the critical exponent is given by $\frac{\xi_1(d)}{\xi_0(d)}$.
By choosing asymptotic form of $D$ at the critical point such that $\nu\sim 1/2 (d\rightarrow \infty)$,we obtained
\begin{equation}
\nu=\frac{-3}{2}+\frac{2\zeta( d/2+c_l)}{\zeta( d/2+c_l+1)}
\end{equation}
To get formula such that lower critical dimension equal to two, we need to set $c_l=0$.
\begin{equation}\label{eq:nuc=0}
\nu=\frac{-3}{2}+\frac{2\zeta(d/2)}{\zeta(d/2+1)}
\end{equation}
To keep correct asymptotic behavior at lower critical dimension($(d-2)\nu\sim1(d\rightarrow 2)$), and $\nu\sim1/2(d\rightarrow \infty)$,natural choice of functional form is perturbation series about d in high dimension,
\begin{equation}
\nu=-\frac{3}{2}+\frac{2\zeta(d/2+f/d+g/d^2+h/d^3+k/d^4+c_l)}{\zeta(d/2+f/d+g/d^2+h/d^3+k/d^4+c_l+1)}
\end{equation}
To get correct asymptotic behavior at lower critical dimension$d_l=2$ requires
\begin{align}
k&=-8f - 4g - 2h - 16c_l \nonumber \\
h&=-12f-4g-32c_l+11.45
\end{align}
Weighted fitting gives
\begin{equation}\label{eq:nufin}
\nu=-\frac{3}{2} + \frac{2\zeta(d/2 - 26.51/d + 156.1/d^2-333.8/d^3+235.7/d^4 +1.224)}{\zeta(d/2 - 26.51/d + 156.1/d^2-333.8/d^3+235.7/d^4 +2.224)}
\end{equation}
For the sake of simplicity, it is written with 4 significant digits.

\section{The estimated value of  the critical exponent $\nu$ and comparison with previous studies}
Expansion series from lower critical dimension is obtained by S.Hikami\cite{hikami92},
\begin{equation}\label{eq:nuH}
\nu\sim \frac{1}{d-2}-\frac{9\zeta(3)}{4}(d-2)^2+\frac{27}{16}\zeta(4)(d-2)^3
\end{equation}

Fitting curve from high dimension was given by \cite{Tarquini2017}
\begin{equation}\label{eq:nuT}
\nu=\frac{1}{2-4.75/d}
\end{equation}

Semi classical theory of the Anderson transition\cite{garcia08} gives
\begin{equation}
\nu=\frac{1}{2}+\frac{1}{d-2}
\end{equation}

For integer dimensions, the estimated value of the critical exponents are listed in Table.\ref{table:integer d}
\begin{table}[ht]
	\begin{center}
	\begin{tabular}{|c||c|c||c||c|} \hline
      	$d$ & $Eq.(\ref{eq:nuc=0})$&$Eq.(\ref{eq:nufin})$& numerical estimate\\ \hline
      	$3$ & $2.395$& $1.569$ &$1.571\pm.004$\cite{slevin14} \\ \hline
      	$4$ & $1.237$ & $1.184$ &$1.156 \pm .014$\cite{ueoka14} \\ \hline
      	$5$ & $0.881$ &  $0.949$ &$0.969 \pm .015$\cite{ueoka14} \\ \hline
      	$6$ & $0.721$ & $0.781$ &$0.78 \pm .06$\cite{Garcia07} \\ \hline
	\end{tabular}
	\end{center}
	\caption{
	Estimated critical exponents for the orthogonal symmetry class for $d=3,4,5$ and $6$ obtained from Eqns.(\ref{eq:nuc=0}), and (\ref{eq:nufin}).
} \label{table:integer d}
\end{table}

For non-integer dimensions, the estimated value of the critical exponents are listed in Table.\ref{table:nu fractal}

\begin{table}[ht]
	\begin{center}
	\begin{tabular}{|l||c|c||c|c||c|} \hline
      	$d$  & $Eq.(\ref{eq:nuc=0})$ & $Eq.(\ref{eq:nufin})$& Refs.\cite{schreiber96,song97,travenec02} \\ \hline
      	$2.22$ & $10.96$ &$4.631$& $4.402\pm.18$  \\ \hline
      	$2.226$ & $10.67$&$4.517$& $2.82\pm.05$ \\ \hline
      	$2.32$ & $7.517$ &$3.317$& $2.59\pm.19$ \\ \hline
      	$2.33$ & $7.287$ &$3.231$& $2.92\pm.14$ \\ \hline
      	$2.365$ & $6.583$ &$2.972$& $2.27\pm.06$ \\ \hline
      	$2.41$ & $5.855$ &$2.709$& $2.50\pm.21$ \\ \hline
      	$2.54$ & $4.436$ &$2.215$& $2.24\pm.31$ \\ \hline
	\end{tabular}
	\end{center}
	\caption{
Same as for Table \ref{table:integer d} but for fractals with spectral dimension
$2<d<3$.
Values of critical exponents in Ref. \cite{schreiber96} were provided by M. Schreiber.
} \label{table:nu fractal}
\end{table}

Within perturbation method, these estimated values are best values without summation method\cite{hikami92}. 

\begin{figure}[th]
\begin{center}
\includegraphics[bb=20 118 575 673,scale=0.4]{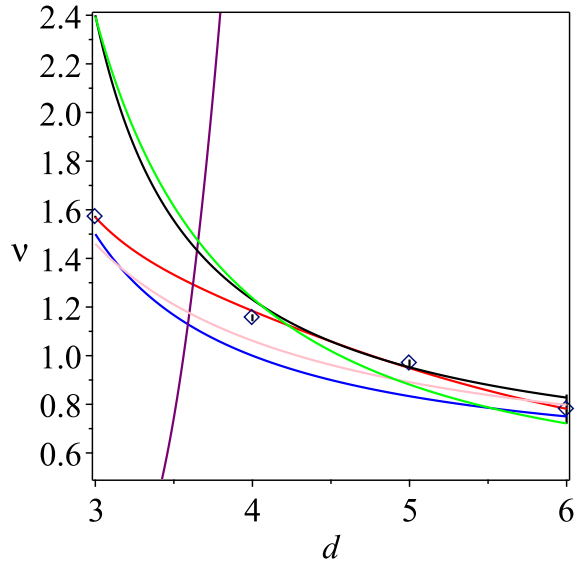}
\vskip13mm
\caption{Comparison between approximation formulas of the critical exponent and numerical estimates in integer dimension.  Red curve is Eq.(\ref{eq:nufin}).Blue curve is semi classical theory of the Anderson transition\cite{garcia08}.Green curve is Eq.(\ref{eq:nuc=0}),Purple curve is Eq.(\ref{eq:nuH}), Black solid curve is Eq.(\ref{eq:nuT}).Pink curve is my previous study\cite{ueoka14}.}
\label{plot:nuint}
\end{center}
\end{figure}

\begin{figure}[th]
\begin{center}
\includegraphics[bb=20 118 575 673,scale=0.4]{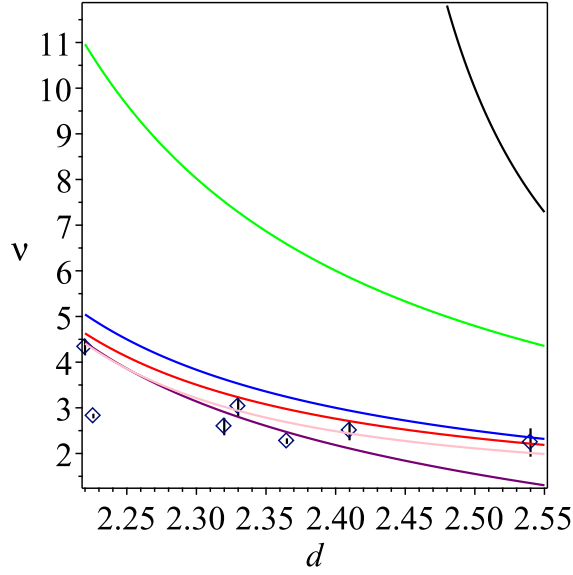}
\vskip13mm
\caption{Comparison between approximation formulas of the critical exponent and numerical estimates in spectral dimension. Red curve is Eq.(\ref{eq:nufin}). Blue curve is semi classical theory of the Anderson transition\cite{garcia08}.Green curve is Eq.(\ref{eq:nuc=0}),Purple curve is Eq.(\ref{eq:nuH}), Black solid curve is Eq.(\ref{eq:nuT}).Pink curve is my previous study\cite{ueoka14}.}
\label{plot:nu}
\end{center}
\end{figure}

In figure \ref{plot:nu} and figure \ref{plot:nuint},numerical estimates and theoretical estimates without summation method are plotted.

Eq.(\ref{eq:nuc=0}) gives almost same estimate with fitting formula from high dimension.\cite{Tarquini2017}.
Fitting formula does not give correct lower critical dimension but  Eq.(\ref{eq:nuc=0}) gives correct lower critical dimension $d_l=2$.
Eq.(\ref{eq:nufin}) gives best estimate over wide range of dimensionality and gives correct asymptotic behavior at $d=2$ and $\nu\sim 1/2(d\rightarrow \infty)$.
Eq.(\ref{eq:nuH}) gives good estimate only near $d=2$.

\section{Conclusion}
Using integral,I take an approximate summation of typical critical Green function in SRWS theory for the Anderson transition in the orthogonal symmetry class.
From obtained perturbation series by lowest degree, we get estimates for the simplest case  which are approximations from upper critical dimension.
They are better than other previous theoretical studies without summation method about dimensionality or fitting formula.
I confirmed that even lowest order estimate gives good estimate of the critical exponent to some extent.
However, I also noticed Pad{\'e} approximant with higher order terms does not work due to existence of pole.
Therefore, suitable summation method which is unknown now is necessary to improve estimate.
This is necessary to improve estimate of the critical exponent without introducing additional degree of freedom by high dimensional approximation.
Thus, problem of estimate of the critical exponent seems to be replaced by finding suitable summation method.

\bibliography{references.bib}

\end{document}